\documentclass[aps,prd,twocolumn,superscriptaddress]{revtex4-2}

\usepackage{amsmath}
\usepackage{amssymb}
\usepackage{slashed}
\usepackage[T1]{fontenc}
\usepackage{hyperref}
\usepackage{graphicx}
\expandafter\ifx\csname package@font\endcsname\relax\else
 \expandafter\expandafter
 \expandafter\usepackage
 \expandafter\expandafter
 \expandafter{\csname package@font\endcsname}%
\fi
\begin{document}

\title{Object-relative ultraviolet weighting of electromagnetic modes and one-loop ultraviolet finiteness of internal photon lines in quantum electrodynamics}%

\author{Christian Rembe}
\email{christian.rembe@tu-clausthal.de}
\affiliation{Clausthal Technical University, Institute of Electrical Information Technology, Leibnizstr. 28\\38678 Clausthal-Zellerfeld, Germany}%

\date{October 2026}%

\begin{abstract}
This work explores whether localized electromagnetic interactions can be modeled by a physically concluded object-relative ultraviolet weighting of photon modes. The proposal is motivated heuristically by a weak-field self-backreaction estimate for localized energy-carrying modes and by a three-dimensional overlap argument. In the resulting ansatz, the infrared sector remains unchanged up to a characteristic angular wavenumber $k_c$, while ultraviolet contributions are suppressed asymptotically by a factor of order $k_c^3/k^3$. The mode structure is object-relative: a real photon is weighted by its frequency $u_\mu k^\mu$ in the rest frame of the interaction object with four-velocity $u^\mu$. The occupation of the modes and the amplitudes of the vacuum fluctuations, in contrast, follow the rules of special relativity, so that virtual photons in loops are weighted by the invariant $k^\mu k_\mu$, i.e.\ by the Euclidean photon momentum. The weighting applies to photon lines only; applied in the same way to the electron line, it conflicts with gauge invariance or with the measured running of the fine-structure constant. One-loop diagrams with internal photon lines become ultraviolet finite, and the weighting can be represented by a gauge-invariant kinetic term of the photon field, so that the Ward--Takahashi identities are preserved. Four test cases are discussed. A Bethe-type estimate of the Lamb shift places the characteristic scale near the electron Compton wavenumber. The anomalous magnetic moment and the relativistic part of the Lamb shift show, however, that a photon weighting at this scale requires additional structure on the electron line to remain compatible with high-precision measurements. This structure, the vacuum polarization, and the spectral consistency of the weighting define the tasks for future work.
\end{abstract}
\maketitle

\section{Introduction}

The Lamb shift was an important experimental result \cite{Lamb1947} that could not initially be explained by straightforward relativistic corrections to the energy levels of the hydrogen atom \cite{Dirac1958}. Even before the experimental discovery of the Lamb shift, Wei\ss kopf \cite{Weisskopf1934} had investigated the electron self-energy, thereby laying the groundwork for understanding how radiative corrections affect atomic energy levels in quantum electrodynamics (QED). Building on this line of thought, Bethe recognized that the shift can be estimated with good accuracy by integrating over photon frequencies up to an upper cutoff at the electron rest energy, with the atomic excitation energies regulating the infrared \cite{Bethe1947}. In order to obtain a finite result, however, he had to subtract the self-energy of a free electron, which is itself divergent. This subtraction is commonly regarded as the birth of renormalization. In particular, the ultraviolet (UV) part of the theory was later brought under systematic control and enabled accurate relativistic calculations of the Lamb shift \cite{KundL1949}. Renormalization became essential for the development of a finite and predictive theory of QED \cite{Tomonaga1946,Schwinger1948a,Feynman1949a,PundV1949,Feynman1949b,Dyson1949,Dyson1950,Feynman1950,Salam1951}, even though it was criticized by pioneers such as Dirac \cite{Dirac1978} and Feynman \cite{Feynman1985}. Nevertheless, renormalized QED has been remarkably successful in reproducing physically important observables. Standard references are Bjorken and Drell \cite{BundD1964}, K\"opp and Kr\"uger \cite{KundK1997}, and Ryder \cite{Ryder1995}.

The necessity of renormalization is often taken to indicate that QED, and perhaps other quantum field theories of the Standard Model, should be understood as effective rather than fundamentally complete descriptions. This motivates the question of whether the ultraviolet structure of the theory may reflect an incomplete physical assumption about the effective contribution of high-frequency modes. The present work is motivated by the idea that a localized field mode should not be treated as spectrally inert with respect to the conditions defining that mode. Since a mode carries energy, and energy gravitates, it is natural to ask whether the corresponding gravitational self-feedback can influence the effective mode structure. As a first step, this possibility is explored here in a weak-field approximation based on Newtonian gravity. Although such an estimate is clearly not adequate near the Planck regime, it provides a simple heuristic starting point and suggests that the standard ultraviolet mode counting may overestimate the effective contribution of sufficiently localized high-frequency modes.

A second line of motivation comes from the role of localization in Bethe's nonrelativistic Lamb-shift calculation. In modern terms, Bethe's treatment already captures the low-energy part of the bound-state self-energy, namely the part that probes the atomic transition structure through virtual dipole couplings to intermediate states. In that sense, it is not merely a historical approximation, but a physically meaningful component of the full Lamb-shift problem. In Bethe's original calculation, however, no distinction is made between situations in which the mode wavelength is large or small compared to the characteristic interaction scale. In three spatial dimensions, the overlap of a mode with a localized interaction region suggests an effective enhancement of coupling efficiency proportional to $k^3$ as long as the wavelength remains larger than the characteristic localization scale. An effective thinning of the mode density by a factor of order $k^{-3}$ would compensate this enhancement and leave the infrared range unchanged up to a characteristic scale. When applied to Bethe's Lamb-shift estimate, the integral becomes UV finite and the characteristic scale is found near the Compton scale.

The aim of the present paper is to explore whether such a modified mode structure can be implemented consistently in QED at one-loop level on photon lines. In particular, Lorentz consistency, ultraviolet finiteness of selected loop integrals, and Ward consistency are examined. As first relativistic test cases, the anomalous magnetic moment and the relativistic part of the Lamb shift are discussed; they reveal what a photon weighting at the Compton scale requires from the electron line and define tasks for future research. The resulting picture is not yet a completed replacement of renormalized QED, but rather a structured first attempt to study whether part of the ultraviolet problem can be reformulated in terms of an effective object-relative weighting of electromagnetic modes. For clarity, Euclidean and Minkowski quantities are distinguished by separate notation, and SI units are used throughout, with $\hbar$ and $c$ kept explicit.

\section{Motivation for and introduction of an effective mode structure for localized interactions}
\label{sec:motivation}

The standard mode density of the electromagnetic field \cite{Demtroeder2015} \cite{Born2013} is obtained by counting stationary modes in a large auxiliary volume and then passing to the continuum limit. In a finite cubic volume $V=L^{3}$ with periodic boundary conditions, the allowed wave vectors $\vec{k}$ are
\begin{equation}
\mathbf{k}=\frac{2\pi}{L}\,\mathbf{n},
\qquad
\mathbf{n}\in\mathbb{Z}^{3},
\end{equation}
so that in the limit $V\to\infty$ one has
\begin{equation}
\sum_{\mathbf{k}}
\;\rightarrow\;
\frac{V}{(2\pi)^{3}}
\int d^{3}k,
\end{equation}
and hence the standard momentum-space mode density
\begin{equation}
dN_{0}
=
\frac{V}{(2\pi)^{3}}\,d^{3}k.
\end{equation}
This construction is highly successful in ordinary quantum electrodynamics. The question addressed in the present work is not whether it should be abandoned in general, but whether it remains fully adequate for sufficiently localized interactions at very high mode frequency.

The proposal studied here is motivated by two independent heuristic considerations. Neither of them is presented as a derivation of the effective weighting used later. Rather, both suggest that sufficiently localized high-frequency electromagnetic modes need not remain completely spectrally unaffected with respect to the physical conditions relevant to the interaction.

\subsection{Heuristic weak-field self-backreaction estimate}

As a first motivation, consider a localized electromagnetic mode in its rest frame and associate to it a characteristic localization scale of order of its wavelength. As a simple isotropic localization model, one may represent such a mode by a spherical resonator whose radius is of order $\lambda$. This construction is not meant to define an exact field-theoretic state. Its role is only to provide a weak-field estimate for the possible size and frequency dependence of gravitational self-backreaction.

A mode of angular frequency $\omega$ carries the energy
\begin{equation}
E=\hbar\omega .
\end{equation}
If this energy is assigned an effective gravitational mass
\begin{equation}
m_{\mathrm{eff}}=\frac{E}{c^{2}}=\frac{\hbar\omega}{c^{2}},
\end{equation}
then a weak-field gravitational potential of order
\begin{equation}
\Phi_{\mathrm{eff}}
\sim
-\frac{Gm_{\mathrm{eff}}}{R}
=
-\frac{G\hbar\omega}{c^{2}R}
\end{equation}
is associated with the localization scale $R$.

In a weak gravitational field, local clock rates and therefore local frequencies are redshifted by a factor $1+\Phi/c^{2}$ to leading order. This suggests a relative frequency correction of order
\begin{equation}
\frac{\delta\omega}{\omega}
\sim
\frac{\Phi_{\mathrm{eff}}}{c^{2}}
\sim
-\frac{G\hbar\omega}{c^{4}R}.
\end{equation}
If the localization scale is taken to be of order the wavelength,
\begin{equation}
R\sim \lambda \sim \frac{c}{\omega},
\end{equation}
one obtains the parametric estimate
\begin{equation}
\delta\omega
\sim
-\frac{G\hbar}{c^{5}}\omega^{3}
=
-\frac{\omega^{3}}{\omega_{P}^{2}},
\end{equation}
up to factors of order unity, with the Planck angular frequency
\begin{equation}
\omega_{P}=\sqrt{\frac{c^{5}}{\hbar G}}.
\end{equation}

Note that, on the Planck scale, the angular frequency spacing is proportional to the preceding frequency. The significance of this estimate is not its precise numerical prefactor, which depends on the chosen localization model, but its qualitative implication: once the energy of a sufficiently localized mode is itself treated as a gravitational source, the resulting self-backreaction need not remain spectrally negligible at arbitrarily high frequency. In particular, the estimate suggests a rapidly increasing frequency-dependent correction. The present work does not derive the later effective ultraviolet weighting from this weak-field argument alone. Rather, the estimate serves only as a first indication that the standard ultraviolet mode counting may overestimate the effective contribution of sufficiently localized high-frequency modes.

Both this estimate and the mode counting above refer to modes that propagate with the phase velocity $c$. The resulting mode weighting for photons cannot, therefore, be transferred directly to particles with rest mass. Throughout this work, the effective mode weighting is applied only to photon lines; fermion propagators are left unchanged, and their proper treatment is an open problem (Sec.~\ref{sec:open}).

\subsection{Heuristic three-dimensional overlap argument}

A second and independent motivation comes from the role of localization in electromagnetic interactions themselves. Consider a localized interaction region characterized by a length scale $l_{c}$ and a corresponding characteristic wave number
\begin{equation}
k_{c}=\frac{2\pi}{l_{c}}.
\end{equation}
In three spatial dimensions, a mode of wavelength $\lambda\sim 1/k$ can heuristically resolve a localized interaction region into a number of spatial subvolumes that grows like $k^{3}$ as long as the wavelength remains larger than the localization scale. Equivalently, the geometric interaction efficiency of the mode with the localized region may be expected to increase proportionally to $k^{3}$ up to the crossover scale $k_{c}$.

If one wishes the infrared effectiveness of the theory to remain unchanged below that scale, a natural compensating ansatz is to thin the ultraviolet mode contribution by the inverse factor. This motivates the introduction of an effective weighting profile
\begin{equation}
W(k;k_{c})=
\begin{cases}
1, & 0\le k\le k_{c},\\
\dfrac{k_{c}^{3}}{k^{3}}, & k>k_{c}.
\end{cases}
\label{eq:W_piecewise}
\end{equation}
The effective interaction-relevant mode density is then modeled as
\begin{equation}
\rho_{\mathrm{eff}}(k;k_{c})
=
\rho_{0}(k)\,W(k;k_{c}),
\end{equation}
where $\rho_{0}(k)$ denotes the standard mode density.

This weighting is introduced here as an effective ansatz, not as a derived law. Its purpose is to encode, in the simplest possible way, the idea that the contribution of sufficiently short-wavelength modes to a localized interaction may be reduced relative to the standard ultraviolet counting, while the infrared sector remains unchanged up to the characteristic scale.

\subsection{Interpretation}

The two heuristic considerations above play different roles. The weak-field self-backreaction estimate suggests that sufficiently localized high-frequency modes need not remain completely unchanged in their spectral role once their own energy is taken seriously as a gravitational source. The three-dimensional overlap argument, by contrast, motivates the specific asymptotic form of the effective weighting used in the present work. The latter is therefore not claimed to follow quantitatively from the former.

The aim of the following sections is not to derive this effective mode structure from a complete underlying theory, but to investigate whether such an object-relative weighting can be implemented in a Lorentz-consistent way and whether it yields a framework in which selected QED amplitudes become ultraviolet finite.

\section{Lorentz consistency of the effective mode structure}
\label{sec:lorentz}

The proposal studied here is formulated as an interaction-relative effective mode structure associated with localized processes. The focus is therefore not on a complete nonlocal field-theoretic embedding, but on the more limited question whether the proposed crossover can be implemented in a Lorentz-consistent way. In particular, the issue is not whether the bare vacuum mode density has the same numerical form in every inertial frame as a function of the same laboratory three-wave number, but whether the same physical crossover is obtained once the electromagnetic mode is described relative to the localized interaction object.

A change of inertial frame reshuffles the electromagnetic mode spectrum through the relativistic Doppler effect. Accordingly, neither the laboratory frequency $\omega$ nor the laboratory wave vector $\mathbf{k}$ of a mode is invariant under Lorentz transformations. This is no more surprising than the fact that the mode spectrum of a resonator is not described by the same numerical frequencies in every inertial frame. What must remain the same is instead the interaction-relative effective mode structure associated with the localized object.

The central conceptual distinction is therefore between

\begin{enumerate}
\item the invariant characteristic scale $k_c$ associated with the localized interaction object, and
\item the covariant mode variable that compares a given electromagnetic mode to that scale.
\end{enumerate}

For an electron, the characteristic scale is naturally expected to lie near the Compton scale,
\begin{equation}
\begin{aligned}
&\lambda_{c} \sim \lambda_C = \frac{h}{m c} \approx 2.426 \times 10^{-12}\, \mathrm{m} \; , \\& \;k_c \sim k_C = \frac{mc}{\hbar} \approx 2.590 \times 10^{12}\, \mathrm{m}^{-1},
\end{aligned}
\end{equation}
or near some effective scale shifted relative to this naive estimate by a geometric factor of order unity.

\subsection{Object-relative mode variable}

Let $u^\mu$ denote the four-velocity of the localized interaction object and let the electromagnetic mode be described by the four-wave vector
\begin{equation}
k^\mu=\left(\frac{\omega}{c},\mathbf{k}\right).
\end{equation}
The natural Lorentz scalar formed from these two quantities is
\begin{equation}
u_\mu k^\mu.
\label{eq:u_dot_k_main}
\end{equation}
This scalar represents the mode frequency measured in the rest frame of the interaction object. To see this explicitly, consider that in the rest frame one has
\begin{equation}
u^\mu=(c,0,0,0),
\end{equation}
and with the metric signature $(+,-,-,-)$ used throughout this work therefore
\begin{equation}
u_\mu=(c,0,0,0).
\end{equation}
It follows that
\begin{equation}
u_\mu k^\mu
=
c\,\frac{\omega}{c}
=
\omega
\end{equation}
in the rest frame of the localized interaction object.

Under a Lorentz transformation,
\begin{equation}
u^\mu \to u'^\mu=\Lambda^\mu{}_\nu u^\nu,
\qquad
k^\mu \to k'^\mu=\Lambda^\mu{}_\nu k^\nu,
\end{equation}
the scalar product is preserved,
\begin{equation}
u'_\mu k'^\mu=u_\mu k^\mu.
\label{eq:u_dot_k_invariant}
\end{equation}
Thus, although the laboratory frequency and propagation direction of a mode are frame dependent, the object-relative mode variable $u\cdot k$ is Lorentz invariant.

The effective crossover should therefore not be formulated as a universal function of the bare laboratory quantity $|\mathbf{k}|/k_c$, but as a function of the invariant ratio
\begin{equation}
\frac{u\cdot k}{c\,k_c}.
\end{equation}
A universal crossover profile may then be written in the form
\begin{equation}
W\!\left(\frac{u\cdot k}{c\,k_c}\right),
\label{eq:W_uk_kc_main}
\end{equation}
with the same function $W$ used in every inertial frame. In the rest frame of the interaction object, this reduces to the expected form
\begin{equation}
W\!\left(\frac{\omega}{c\, k_c}\right),
\end{equation}
or, for a real electromagnetic mode with $\omega=c|\mathbf{k}|$,
\begin{equation}
W\!\left(\frac{|\mathbf{k}|}{k_c}\right).
\end{equation}
If the ultraviolet asymptotics satisfy
\begin{equation}
W(x)\sim x^{-3},
\qquad x\gg 1,
\end{equation}
then in the rest frame one recovers the intended asymptotic thinning
\begin{equation}
W \sim \left(\frac{k_c}{|\mathbf{k}|}\right)^3 .
\end{equation}

In three spatial dimensions, the mode direction generally changes under a boost, and the observed Doppler shift depends on the angle between the propagation direction and the velocity of the interaction object. This dependence is already encoded in the scalar $u\cdot k$. Explicitly,
\begin{equation}
u_\mu k^\mu
=
\gamma\omega
-
\gamma\,\mathbf{v}\cdot\mathbf{k},
\label{eq:u_dot_k_explicit_main}
\end{equation}
with
\begin{equation}
u^\mu=\gamma(c,\mathbf{v}),
\qquad
\gamma=\frac{1}{\sqrt{1-v^2/c^2}}.
\end{equation}
Thus the use of $u\cdot k$ automatically incorporates the full three-dimensional angular dependence of the boost.

For real electromagnetic modes, a characterization solely in terms of $k^\mu k_\mu$ is not well suited to the present mode-based motivation. On the light cone one has $k^\mu k_\mu=0$, so that this invariant alone does not distinguish low and high object-relative mode frequency. For real photons, the physically relevant crossover is therefore expressed in terms of the invariant object-relative variable $u\cdot k$ together with the invariant characteristic scale $k_c$.

The resulting Lorentz-consistency statement may therefore be summarized as follows:

\begin{quote}
There exists a universal interaction-relative effective mode structure such that, for every localized particle or interaction region, the same effective crossover profile is realized in every inertial frame once the modes are described relative to the boosted object and to its invariant characteristic scale $k_c$.
\end{quote}

\subsection{Mode structure and fluctuation amplitudes}
\label{sec:hypothesis}

The variable $u\cdot k$ refers to real photons on the light cone. Loop diagrams, however, contain virtual photons, which are not confined to the light cone, and require a separate statement. It is obtained by distinguishing two aspects of the photon field:

\begin{quote}
\textit{Hypothesis.} The mode structure, i.e.\ the spectrum of available modes, is object-relative. Since every inertial system forms its own causal resonator~\cite{Rembe2026}, the possible mode spectrum is observer related, and a real photon is weighted by its frequency $u\cdot k$ relative to the object. The occupation of the modes and the amplitudes of the vacuum fluctuations, on the other hand, follow the standard rules of special relativity; the object enters their weighting only through the scale $k_c$.
\end{quote}

The two variables therefore refer to different physical situations. A real photon is a persistent quantum on the light cone, where $k^{2}=0$ carries no information; what distinguishes it is its frequency relative to the object, $u\cdot k$, which determines the resonator mode it occupies. A virtual photon is a short-lived fluctuation without a definite occupation. It is not confined to the light cone, and its deviation from it, the virtuality $k^{2}$, is limited only by the energy--time uncertainty relation. Its weighting is therefore a function of the invariant $k^{2}$. After Wick rotation to Euclidean momentum space \cite{Wick1954}, $-k^{2}$ becomes $K_{E}^{2}$, so that internal photon lines are weighted by the magnitude $K_{E}$ of the Euclidean photon momentum,
\begin{equation}
W=W(K_{E};k_{c}),
\label{eq:W_KE}
\end{equation}
with the object entering only through $k_{c}$.

This choice is also forced by the structure of loop integrals. A weighting of the frequency alone, $u\cdot k$, would after Wick rotation act on a single Euclidean component of the photon momentum, and a weighting of the spatial momentum $|\mathbf{k}|$ in the rest frame does not decouple: for the anomalous magnetic moment of Sec.~\ref{sec:formfactors} it changes the result by $13$ to $33\,\%$ even for $k_{c}$ between $k_{C}$ and $10\,k_{C}$, in line with the known non-decoupling of Lorentz-violating regulators \cite{Collins2004}. Since the piecewise profile of Eq.~\eqref{eq:W_piecewise} is not analytic, the weighting of internal lines is defined directly in Euclidean space; whether an equivalent Minkowski-space form exists is left open. All loop results of this work refer to Eq.~\eqref{eq:W_KE}.

These statements should be understood in a deliberately limited sense. They do not yet establish the consistency of the proposal with the full spectral and analytic structure expected of a relativistic quantum field theory. In particular, consistency with a K\"allen--Lehmann-type representation \cite{Kaellen1952,Lehmann1954},
\begin{equation}
D(p^2)
=
\int_0^\infty d\mu^2\,
\frac{\rho(\mu^2)}{p^2-\mu^2+i\epsilon},
\end{equation}
with the spectral density $\rho(\mu^2)$ and the spectral mass parameter $\mu$, is an important benchmark. It is taken up in Sec.~\ref{sec:action}.

\section{Ward Consistency of the Effective Mode Structure}
\label{sec:ward}

A central consistency requirement of quantum electrodynamics is the Ward identity \cite{Ward1950},
which relates the electron self-energy to the vertex correction. In momentum space it reads
\begin{equation}
q_\mu \Gamma^\mu(p+q,p)=S^{-1}(p+q)-S^{-1}(p),
\label{eq:WTI}
\end{equation}
where $S^{-1}(p)$ denotes the full inverse electron propagator and $\Gamma^\mu(p+q,p)$ the full one-particle-irreducible vertex function.

In the present framework, the effective mode structure is not introduced as a modification of the elementary QED vertex. It changes the mode content of the photon field, from which the internal photon lines are built, and is interpreted as a scalar weighting of internal electromagnetic modes relative to the localized interacting particle.
Accordingly, the relevant consistency question is whether such a weighting can be implemented without spoiling the Ward relation between self-energy and vertex corrections.

\subsection{One-loop check}

At one-loop level, let the electron self-energy be written schematically as
\begin{equation}
\begin{aligned}
c\Sigma(p)=&(-ie)^2 c \int \frac{d^4k}{(2\pi)^4}\,
W(k;u,k_c)\,\\&
\gamma^\nu\,S_F(p-\hbar k)\,\gamma^\mu\,D_{\mu\nu}(k),
\label{eq:Sigma_weighted}
\end{aligned}
\end{equation}
where $D_{\mu\nu}(k)$ is the standard photon propagator, $S_F$ the standard fermion propagator, and $W(k;u,k_c)$ the effective scalar mode weighting.
For internal photon lines, $W$ depends on the invariant $k^{2}$ of the virtual photon, with the scale $k_{c}$ set by the interacting particle (Sec.~\ref{sec:hypothesis}); the notation $W(k;u,k_c)$ is kept here, since the argument below holds for any scalar weighting of the photon mode.

The corresponding one-loop vertex correction is
\begin{equation}
\begin{aligned}
\Lambda^\rho&(p+q,p)=(-ie)^2 \int \frac{d^4k}{(2\pi)^4}\,
W(k;u,k_c)\,
\gamma^\nu\,\\
&S_F(p+q-\hbar k)\,\gamma^\rho \,S_F(p-\hbar k)\,\gamma^\mu\,D_{\mu\nu}(k).
\label{eq:Vertex_weighted}
\end{aligned}
\end{equation}
The crucial requirement is that the \emph{same} scalar function $W(k;u,k_c)$ multiplies the same internal photon momentum $k$ in both Eqs.~\eqref{eq:Sigma_weighted} and \eqref{eq:Vertex_weighted}.

Contracting Eq.~\eqref{eq:Vertex_weighted} with the external photon momentum $q_\rho$ and using the standard fermion-line identity
\begin{equation}
q_\rho \gamma^\rho = S_0^{-1}(p+q- \hbar k)-S_0^{-1}(p- \hbar k),
\label{eq:fermion_line_identity}
\end{equation}
with the free inverse propagator
\begin{equation}
S_0^{-1}(r)=\slashed{r}-m \,c,
\label{eq:free_inverse_propagator}
\end{equation}
one obtains
\begin{equation}
\begin{aligned}
q_\rho\,S_F(p+q-\hbar k)\,\gamma^\rho\,S_F(p-\hbar k)
=&
S_F(p-\hbar k)-\\&S_F(p+q-\hbar k).
\label{eq:telescope_identity}
\end{aligned}
\end{equation}
Since the factor $W(k;u,k_c)$ is a common scalar multiplier depending only on the internal photon mode $k$, it is carried through this contraction unchanged.
Equation~\eqref{eq:Vertex_weighted} therefore gives
\begin{equation}
\begin{aligned}
q_\rho \Lambda^\rho&(p+q,p)
=
(-ie)^2 \int \frac{d^4k}{(2\pi)^4}\,
W(k;u,k_c)\,
\gamma^\nu \\
&\Bigl[S_F(p-\hbar k)-S_F(p+q-\hbar k)\Bigr]
\gamma^\mu D_{\mu\nu}(k).
\end{aligned}
\label{eq:vertex_contracted}
\end{equation}
Comparing with Eq.~\eqref{eq:Sigma_weighted}, one finds
\begin{equation}
q_\rho \Lambda^\rho(p+q,p)=\Sigma(p)-\Sigma(p+q),
\label{eq:weighted_WTI_loop}
\end{equation}
up to the usual overall sign conventions.
Including the tree-level vertex, one recovers the standard one-loop Ward relation in the form
\begin{equation}
q_\mu \Gamma^\mu(p+q,p)=S^{-1}(p+q)-S^{-1}(p).
\label{eq:weighted_WTI_full}
\end{equation}

Thus, at one-loop level the Ward identity remains compatible with the effective mode structure provided that the weighting enters as the same scalar factor associated with the same internal photon momentum in both the self-energy and the vertex correction.

\subsection{More general structural statement}

The one-loop check above reflects a more general structural point.
Suppose that in a class of abelian loop diagrams each internal photon line with momentum $k$ is multiplied by a scalar factor
\begin{equation}
D_{\mu\nu}(k)\;\longrightarrow\;W(k;u,k_c)\,D_{\mu\nu}(k),
\label{eq:weighted_propagator_general}
\end{equation}
where $W(k;u,k_c)$ depends only on the photon mode and on fixed external characteristics of the interacting particle, such as $u^\mu$ and $k_c$, but carries no additional Lorentz, Dirac, or gauge structure.

Then the contraction with the external photon momentum acts only on the fermion line, exactly as in standard QED.
If, for a given loop momentum $k$, the unweighted integrands satisfy a diagrammatic identity of the form
\begin{equation}
q_\mu \mathcal{I}^\mu_\Lambda(p,q;k)
=
\mathcal{I}_\Sigma(p;k)-\mathcal{I}_\Sigma(p+q;k),
\label{eq:diagrammatic_identity}
\end{equation}
then multiplication by the same scalar factor $W(k;u,k_c)$ gives immediately
\begin{equation}
\begin{aligned}
q_\mu \Bigl[W(k;u,k_c)\,&\mathcal{I}^\mu_\Lambda(p,q;k)\Bigr]
=
W(k;u,k_c)\,\\
&\Bigl[\mathcal{I}_\Sigma(p;k)-\mathcal{I}_\Sigma(p+q;k)\Bigr].
\end{aligned}
\label{eq:diagrammatic_identity_weighted}
\end{equation}
This is the diagrammatic argument by which the Ward--Takahashi identity for
the vertex is established in QED; it applies because the contraction with $q$
acts only on charged lines, while the photon lines and their weights are
spectators. After integration over $k$, the same telescoping structure is preserved.
Hence Ward consistency is maintained for the class of implementations in which the effective mode structure enters as a common scalar weighting of internal photon modes.

\subsection{Scope of the diagrammatic argument}

The diagrammatic argument above shows the following specific statement:

\begin{quote}
If the effective mode structure is implemented as a common scalar weighting factor associated with each internal photon mode, and if the same weighting is used consistently in all self-energy and vertex contributions linked by the Ward identity, then the standard one-loop Ward relation is preserved.
\end{quote}

It would fail if the effective mode structure were introduced in a way that modifies the elementary vertex itself, depends differently on external fermion momenta in self-energy and vertex graphs, or assigns different weights to the same internal photon mode in different diagrams. By itself, the diagrammatic argument does not establish the complete hierarchy of Ward--Takahashi identities to all orders \cite{Takahashi1957}. This follows from the formulation at the level of the action given next.

\subsection{Formulation at the level of the action}
\label{sec:action}

At the level of the action, the weighting of the photon line can be represented by a gauge-invariant kinetic term of the photon field,
\begin{equation}
  S_{\gamma}=-\frac{1}{4\mu_{0}}\int\mathrm{d}^{4}x\;
  F_{\mu\nu}\,W^{-1}\bigl(\hat K\bigr)\,F^{\mu\nu},
  \label{act:kinetic}
\end{equation}
where $\hat K$ denotes the operator whose eigenvalue on a plane wave is the magnitude $K_{E}$ of the Euclidean photon momentum after Wick rotation. With the gauge-fixing term weighted in the same way, the Feynman-gauge photon propagator, Eq.~\eqref{eq:photon_propagator} below, is multiplied by $W$, while the vertex is left unchanged. Since the Ward--Takahashi identities of QED follow from current conservation in the fermion sector, they then hold to all orders and not only in the one-loop check above, in analogy with higher-covariant-derivative regularization~\cite{Slavnov1972}.

Equation~\eqref{act:kinetic} is a formulation of the weighting at the level of the action, not a derivation of it; $W$ itself remains motivated by the mode-density considerations of Sec.~\ref{sec:motivation}. Two points remain open. First, a propagator falling faster than $1/k^{2}$ requires a spectral density that is not everywhere positive~\cite{Kaellen1952,Lehmann1954}, and whether this is compatible with unitarity is not established. Second, Eq.~\eqref{act:kinetic} contains a single scale $k_{c}$, which is sufficient as long as only one kind of interacting object is present; how the dependence of $k_{c}$ on the interacting object is expressed at the level of the action is not yet known. The kinetic term describes the weighting of virtual photons; on the light cone it reduces to unity, so that the object-relative weighting of real photons by $u\cdot k$ is not contained in it and belongs to the same open question.

\section{Ultraviolet Behavior of the Weighted One-Loop Integrals}
\label{sec:uv}

The present proposal is formulated in terms of an effective mode structure whose scale is set by the localized interacting particle, rather than by a property of the free vacuum alone.
Accordingly, the ultraviolet question must also be phrased in this interaction-relative sense:
does the effective weighting of the internal electromagnetic modes render the relevant one-loop QED integrals ultraviolet finite?

\subsection{Asymptotic weighting of the internal photon line}

Let $W(K_E;k_c)$ denote the effective scalar weighting assigned to an internal photon line with Euclidean photon momentum $K_E$, Eq.~\eqref{eq:W_KE}, where the interacting particle sets the characteristic scale $k_c$.
The asymptotic assumption of the present approach is that for sufficiently large photon momentum the effective mode structure falls as
\begin{equation}
W \sim \frac{k_c^3}{K_E^3},
\qquad K_E \gg k_c .
\label{eq:mu_rest_asymptotic}
\end{equation}
This is the loop-integral expression of the proposed thinning of the effective electromagnetic mode structure.

The significance of the power $K_E^{-3}$ is that in four dimensions the radial Euclidean integration measure \cite{Wick1954} grows as $K_E^3\,dK_E$.
Hence the asymptotic mode weighting compensates the growth of the radial measure and removes the leading phase-space enhancement that is responsible for the ultraviolet sensitivity of the unweighted one-loop integrals.

\subsection{Euclidean one-loop power counting of the photon line}

After Wick rotation \cite{Wick1954}, the Euclidean integration measure is
\begin{equation}
d^4K_E = 2\pi^2 K_E^3\,dK_E ,
\label{eq:euclidean_measure}
\end{equation}
with $K_E = |K_E^\mu|$ the magnitude of the Euclidean photon momentum, the loop momentum being routed through the photon line.
For a weighted one-loop contribution of the generic form
\begin{equation}
I \sim \int \frac{d^4K_E}{(2\pi)^4}\,
W(K_E)\,
\mathcal{F}(K_E,\text{external momenta}),
\label{eq:generic_weighted_loop}
\end{equation}
the ultraviolet behavior is determined by the product of the radial measure, the effective mode weighting, and the usual propagator structure contained in $\mathcal{F}$.

Using Eq.~\eqref{eq:euclidean_measure} together with the asymptotic form
\begin{equation}
W(K_E)\sim \frac{k_c^3}{K_E^3},
\qquad K_E \gg k_c ,
\label{eq:mu_euclidean_asymptotic}
\end{equation}
one finds
\begin{equation}
K_E^3\,W(K_E)\sim k_c^3 .
\label{eq:measure_compensation}
\end{equation}
Thus the ultraviolet growth of the four-dimensional radial phase-space measure is compensated by the effective mode weighting.
The remaining ultraviolet behavior is then governed only by the standard QED denominators and numerators contained in $\mathcal{F}$.

This observation is sufficient for the present purpose:
for the one-loop QED amplitudes with an internal photon line considered here, the remaining propagator structure provides additional inverse powers of $K_E$, so that the weighted integrals are ultraviolet finite.
In this sense the effective mode structure removes the leading ultraviolet phase-space growth of the loop integrals.

\subsection{Interpretation and limitations}

The ultraviolet finiteness obtained here should be understood as a one-loop power-counting consequence of the proposed interaction-relative effective mode structure.
It can be concluded:
if the internal electromagnetic modes relevant to the interacting particle are weighted asymptotically by a factor of order $k_c^3/K_E^3$, then the leading four-dimensional phase-space growth of the one-loop integrals is removed, and the remaining standard QED propagator structure is sufficient to make the weighted one-loop amplitudes with internal photon lines ultraviolet finite.

Thus, within the present framework, ultraviolet renormalization is replaced at one-loop level, for diagrams with internal photon lines, by an effective interaction-relative mode weighting.
The mass subtraction remains necessary, but it becomes finite (Sec.~\ref{sec:bethe}). The charge renormalization, in contrast, is carried by the vacuum polarization, which contains no internal photon line; it is not treated here and remains as in standard QED (Sec.~\ref{sec:open}).

A full analysis of higher orders, overlapping divergences, and possible extensions beyond abelian QED remains an open problem.

\section{Examples}
\label{sec:examples}

Four examples are discussed. The Casimir effect serves as a consistency test and shows that the weighting leaves the force between conducting plates unchanged. A compact one-loop test shows directly that the weighting renders diagrams with an internal photon line ultraviolet finite. A Bethe-type estimate of the Lamb shift, evaluated without a lower cutoff, fixes the order of magnitude of the characteristic scale. Finally, the form factors of the free electron, which determine the anomalous magnetic moment and the relativistic part of the Lamb shift, serve as a relativistic test case and show what a photon weighting at this scale requires from the electron line. The open questions that result for the electron line are collected at the end.

\subsection{Casimir Effect as a Consistency Test}
\label{sec:casimir}

One of the most important experimental manifestations of vacuum fluctuations
is the Casimir effect \cite{Casimir1948}. The Casimir effect provides a
sensitive test for any modification of the vacuum mode spectrum, since the
observable force originates from the difference between a discrete mode
spectrum and the corresponding continuum limit.

For two perfectly conducting plates at distance $L$, only the wavenumber
component normal to the plates is discrete, $k_{z}=n\pi/L$. After integration
over the transverse wavenumbers and summation over both polarizations, the
energy per unit area can be written as~\cite{Bordag2001}
\begin{equation}
\frac{E(L)}{A}=\frac{\hbar c}{2\pi}
\Biggl[\,\frac12F(0)+\sum_{n=1}^{\infty}F\!\left(\frac{n\pi}{L}\right)
-\frac{L}{\pi}\int_{0}^{\infty}\!d\kappa\,F(\kappa)\Biggr],
\label{cas:energy}
\end{equation}
with
\begin{equation}
F(\kappa)=\int_{\kappa}^{\infty}\!du\,u^{2}\,W(u),
\end{equation}
and $u$ is the modulus of the wavevector. The factor $1/2$ of the term
$n=0$ arises because the mode with $k_{z}=0$ has only one polarization: its
electric field is normal to the plates, whereas a field parallel to ideal
conductors vanishes everywhere. All modes with $n\ge1$ have two polarizations,
which are contained in the prefactor. The modes between the plates are real modes of a resonator, and
the weighting $W$ enters as a function of their frequency $\omega=cu$ in the
rest frame of the plates, in accordance with Sec.~\ref{sec:lorentz}.

With the weighting $W(u)\sim(k_{c}/u)^{3}$ for $u\gg k_{c}$, the integrand
$u^{2}W(u)$ approaches the constant $k_{c}^{3}$, so that $F(\kappa)$ contains
an $L$-independent part that grows linearly with the upper limit. This
reflects the fact that the absolute vacuum energy per volume, whose spectral
density $\rho(k)\,\hbar ck\,W(k)$ also approaches a constant, is not rendered
finite by the object-relative weighting; the absolute vacuum energy is not an
object-relative quantity, and its treatment in terms of the causal resonator
is discussed in Ref.~\cite{Rembe2026}. Being independent of $L$, this part
does not contribute to the force; it is removed by the integral term in
Eq.~\eqref{cas:energy}. For real plates, the spectrum entering
Eq.~\eqref{cas:energy} terminates smoothly at the upper end, since metals
become transparent above their plasma frequency (see below); mathematically,
any smooth regulator has the same effect. The boundary terms of the
Euler--Maclaurin formula at the upper end then vanish, and
\begin{equation}
\begin{split}
&\frac12F(0)+\sum_{n=1}^{\infty}F\!\left(\frac{n\pi}{L}\right)
-\frac{L}{\pi}\int_{0}^{\infty}\!d\kappa\,F(\kappa)\\
&\qquad=-\sum_{j\ge1}\frac{B_{2j}}{(2j)!}\left(\frac{\pi}{L}\right)^{2j-1}
F^{(2j-1)}(0).
\end{split}
\label{cas:EM}
\end{equation}
The half weight of the term $n=0$ is the trapezoidal end correction for which
the Euler--Maclaurin formula holds in this form. The derivatives are evaluated
at the \emph{lower} end of the spectrum. Since $F'(\kappa)=-\kappa^{2}W(\kappa)$,
one has $F'(0)=0$ and $F'''(0)=-2W(0)=-2$, and with $B_{4}=-1/30$
\begin{equation}
\frac{E(L)}{A}=-\frac{\pi^{2}\hbar c}{720\,L^{3}},
\label{cas:result}
\end{equation}
the standard Casimir result. It depends on the weighting only through
$W(0)=1$; higher terms are suppressed as long as $W$ is smooth and
$k_{c}\gg\pi/L$. The linear growth of $F(\kappa)$ at large $\kappa$ does not
enter, since it affects neither the derivatives at $\kappa=0$ nor, with a
smooth termination, the upper end.

Two qualifications follow. First, the smoothness requirement is not merely
technical. For the piecewise profile, the jump of $W'$ at $k_{c}$ produces an
additional term that oscillates with $L$,
\begin{equation}
\frac{\Delta E}{E_{\mathrm{Cas}}}=-180\,x\,\tilde B_{3}(\{x\}),
\qquad x=\frac{k_{c}L}{\pi},
\label{cas:kink}
\end{equation}
where $B_{3}(t)=t^{3}-\tfrac32t^{2}+\tfrac12t$ is the third Bernoulli
polynomial, evaluated at the fractional part $\{x\}$ of $x$, so that
$B_{3}(\{x\})$ is periodic in $x$ with period one. Its amplitude grows linearly
with $k_{c}L$. This is an artifact of the kink combined with ideal
conductors; for the Casimir problem the weighting must be taken as a smooth
profile, e.g.\ $W(x)=(1+x^{a})^{-3/a}$ with $x=k/k_{c}$, for which the
corrections vanish for $k_{c}L\gg\pi$. Second, if the crossover were placed at
the scale of the plate distance, $k_{c}\sim\pi/L$, the modes that generate the
force would themselves be weighted, and the Casimir energy would change by an
amount of order unity that depends on the detailed shape of $W$. Since Casimir
forces are measured at the percent level~\cite{Lamoreaux1997,MohideenRoy1998},
this is excluded. The plate distance sets the scale of the observable, not the
crossover: the Casimir effect requires $k_{c}\gg\pi/L$.

For real plates, the relevant condition is even weaker. Metals are
transparent above their plasma frequency, $\hbar\omega_{p}\approx10\,
\mathrm{eV}$, corresponding to $k_{p}=\omega_{p}/c\approx5\times10^{7}\,
\mathrm{m}^{-1}$. Modes with $k\gg k_{p}$ are the same inside and outside the
plates and do not contribute to the force. A weighting with
$k_{c}\approx k_{C}=2.6\times10^{12}\,\mathrm{m}^{-1}$, as suggested by the
Lamb-shift estimate in Sec.~\ref{sec:bethe}, therefore leaves the Casimir
force unchanged.
The Casimir effect is a consistency test of the weighting, not a determination
of its scale: it requires the crossover to lie far above the modes that
generate the force, and it does not require the individual vacuum
contributions to remain unchanged.

\subsection{Ultraviolet one-loop test}
\label{sec:uvtest}

The purpose of the present example is not to extract a new
observable-dependent characteristic scale, but to illustrate in the most
direct way how the proposed effective mode structure modifies the ultraviolet
part of a one-loop QED contribution with an internal photon line, and to make
the power counting of Sec.~\ref{sec:uv} explicit for the self-energy and the
vertex correction.

Consider a one-loop contribution with one internal photon line after Wick
rotation to Euclidean momentum space. With the loop momentum routed through
the photon line, it has the schematic form
\begin{equation}
I_{\mathrm{1loop}}
\sim
\int \frac{d^{4}K_{E}}{(2\pi)^{4}}\,
W(K_{E};k_{c})\,
F(K_{E},\text{external}),
\end{equation}
where $K_{E}=|K_{E}^{\mu}|$ denotes the Euclidean photon momentum, $F$
summarizes the standard propagator and numerator structure of the
corresponding QED diagram, and $W(K_{E};k_{c})$ is the effective
object-relative mode weighting. Since the weighting is attached to the photon
line, $K_{E}$ must be the photon momentum; a shifted loop variable, as
produced by Feynman parametrization, is not a valid argument of $W$.

For the present framework, the weighting is taken in the piecewise form of
Eq.~\eqref{eq:W_piecewise},
\begin{equation}
W(K_{E};k_{c})=
\begin{cases}
1, & 0\le K_{E}\le k_{c},\\
\dfrac{k_{c}^{3}}{K_{E}^{3}}, & K_{E}>k_{c}.
\end{cases}
\label{eq:W_piecewise_KE}
\end{equation}
The infrared range therefore remains unchanged up to the characteristic scale
$k_{c}$, while the ultraviolet part is suppressed by the factor
$k_{c}^{3}/K_{E}^{3}$.

Using spherical coordinates in four Euclidean dimensions,
\begin{equation}
d^{4}K_{E}=2\pi^{2}K_{E}^{3}\,dK_{E},
\end{equation}
the integral becomes
\begin{equation}
I_{\mathrm{1loop}}
\sim
\frac{1}{8\pi^{2}}
\int_{0}^{\infty}
dK_{E}\,
K_{E}^{3}\,
W(K_{E};k_{c})\,
F(K_{E},\text{external}).
\end{equation}

The ultraviolet effect of the weighting is now immediate. For
$K_{E}\gg k_{c}$,
\begin{equation}
K_{E}^{3}\,W(K_{E};k_{c})
\sim
k_{c}^{3},
\end{equation}
so that the growth of the four-dimensional radial measure is compensated. The
ultraviolet behavior is therefore governed only by the remaining asymptotics
of the standard QED kernel $F$. For large $K_{E}$,
\begin{equation}
F(K_{E},\text{external})\sim \frac{1}{K_{E}^{n}},
\end{equation}
with $n=3$ for the electron self-energy and $n=4$ for the vertex correction,
corresponding to the superficial degrees of divergence one and zero. The
weighted ultraviolet tail behaves as
\begin{equation}
I_{\mathrm{UV}}
\sim
k_{c}^{3}
\int_{k_{c}}^{\infty}
\frac{dK_{E}}{K_{E}^{n}}
=\frac{k_{c}^{4-n}}{n-1},
\end{equation}
which is finite for $n>1$. The borderline case $n=1$ would correspond to a
logarithmic divergence of the weighted integral; the self-energy and the
vertex correction lie well inside the convergent range.

It is useful to introduce the dimensionless variable
\begin{equation}
\kappa=\frac{K_{E}}{k_{c}},
\end{equation}
for which the one-loop contribution takes the form
\begin{equation}
I_{\mathrm{1loop}}
\sim
\frac{k_{c}^{4}}{8\pi^{2}}
\int_{0}^{\infty}
d\kappa\,
\kappa^{3}\,
W(\kappa)\,
F(k_{c}\kappa,\text{external}),
\end{equation}
with
\begin{equation}
W(\kappa)=
\begin{cases}
1, & 0\le \kappa\le 1,\\
\kappa^{-3}, & \kappa>1.
\end{cases}
\end{equation}

This form makes the physical content of the proposal particularly
transparent. The infrared region $\kappa\lesssim 1$ is left unchanged,
whereas the ultraviolet region $\kappa\gtrsim 1$ is thinned by precisely the
factor required to cancel the radial $\kappa^{3}$ growth of the
four-dimensional loop measure. The example shows directly that the weighting
acts above the characteristic scale and renders the ultraviolet part of
one-loop contributions with an internal photon line finite.

For an electron, the characteristic scale $k_{c}$ is naturally expected to lie
near the Compton scale. In this sense, the present example illustrates in
compact form how the ultraviolet sector above that scale is suppressed
without modifying the infrared part of the one-loop structure. The finite
parts that remain are, however, not free: as shown in Sec.~\ref{sec:formfactors},
a weighting at the Compton scale changes the finite form factors $F_{2}(0)$ and
$F_{1}'(0)$ in a way that has to be compensated on the electron line.

\subsection{Low-energy Lamb-shift estimate with modified mode weighting}
\label{sec:bethe}

For the Lamb-shift problem, the electron is part of a bound state in the
Coulomb field and does not define an isolated global rest frame in the same
simple way. Since the electron moves with velocities of order $\alpha c$,
however, its rest frame and that of the atom coincide to leading order. In
Bethe's nonrelativistic treatment~\cite{Bethe1947}, the intermediate photons appear as quanta on the light cone, and their weighting is a function of the
photon angular frequency $\omega$ in this frame (Sec.~\ref{sec:lorentz}). The
level shift after subtraction of the self-energy of the free electron is then,
with the effective mode weighting inserted,
\begin{equation}
  \begin{split}
  \Delta E_{n}={}&\frac{2\alpha}{3\pi m^{2}c^{2}}
  \sum_{m}\bigl|\langle m|\mathbf{p}|n\rangle\bigr|^{2}\,\hbar\omega_{mn}\\
  &\cdot\int_{0}^{\infty}\mathrm{d}\omega\,
  \frac{W(\omega)}{\omega_{mn}+\omega},
  \qquad \omega_{mn}=\frac{E_{m}-E_{n}}{\hbar}.
  \end{split}
  \label{eq:bethe_exact}
\end{equation}
The integral starts at $\omega=0$. The infrared region is regulated by the
denominator $\omega_{mn}+\omega$, i.e.\ by the atomic binding, and requires no
lower cutoff. The sum over intermediate states obeys the sum rule
\begin{equation}
  \begin{split}
  \sum_{m}\bigl|\langle m|\mathbf{p}|n\rangle\bigr|^{2}\,\hbar\omega_{mn}
  &=\frac{\hbar^{2}}{2}\,\langle n|\nabla^{2}V|n\rangle\\
  &=2\pi\hbar^{3}\alpha c\,|\psi_{n}(0)|^{2},
  \end{split}
  \label{eq:sumrule}
\end{equation}
so that in hydrogen only S states contribute at this order, with
$|\psi_{n}(0)|^{2}=1/(\pi n^{3}a_{0}^{3})$. With the weights
$w_{nm}=|\langle m|\mathbf{p}|n\rangle|^{2}\,\hbar\omega_{mn}$, the mean
excitation frequency $\bar\omega_{n}$ is defined by the logarithmic average
\begin{equation}
  \ln\frac{\hbar\bar\omega_{n}}{\mathrm{Ry}}
  =\frac{\sum_{m}w_{nm}\ln\bigl(\hbar|\omega_{mn}|/\mathrm{Ry}\bigr)}
        {\sum_{m}w_{nm}}
  \equiv\ln k_{0}(n),
  \label{eq:bethelog}
\end{equation}
where $\ln k_{0}(n)$ is the Bethe logarithm. For the 2S state,
$\ln k_{0}(2S)=2.811769893$~\cite{DrakeSwainson1990}, hence $\hbar\bar\omega_{2S}=16.64\,\mathrm{Ry}=226.4\,\mathrm{eV}$, or, with
$\mathrm{Ry}=\alpha^{2}mc^{2}/2$,
$\bar k_{2S}=\bar\omega_{2S}/c=\tfrac12\alpha^{2}k_{0}(2S)\,k_{C}
=k_{C}/2257=1.147\times10^{9}\,\mathrm{m}^{-1}$. The mean excitation frequency lies far above the ionization
frequency $\mathrm{Ry}/\hbar$ because the weights in Eq.~\eqref{eq:bethelog}
are dominated by continuum states.

The piecewise profile of Eq.~\eqref{eq:W_piecewise}, written in the photon
angular frequency, provides the weighting
\begin{equation}
  W(\omega)=
  \begin{cases}
    1, & \omega<\omega_{c},\\[2pt]
    (\omega_{c}/\omega)^{3}, & \omega\geq\omega_{c},
  \end{cases}
  \qquad \omega_{c}=c\,k_{c}.
  \label{eq:W_lamb}
\end{equation}
The subtraction of the free-electron self-energy remains necessary on
physical grounds: the free electron carries a self-energy as well, and the
measured mass already contains it, so that observables measure only changes
relative to the free electron. With the weighting, the subtraction is finite,
of order $\alpha/\pi$ relative to $mc^{2}$, and becomes an ordinary
conversion between bare and observed mass rather than the removal of a
divergence. The bare mass $m_{0}=m-\delta m$ is thereby a finite quantity
with its own physical content, which is the prerequisite for the
electron-line structure discussed in Sec.~\ref{sec:open}. For the 2S--2P
splitting the subtraction cancels identically, since
$\langle\mathbf{p}^{2}\rangle$ is the same for both states.

Replacing the individual $\omega_{mn}$ by the mean frequency
$\bar\omega_{n}$, the frequency integral splits at $\omega_{c}$. Below the
crossover, $W=1$ gives a logarithm; above it, the substitution
$t=\omega_{c}/\omega$ turns the weighted tail into an integral over
$0\le t\le1$,
\begin{equation}
  \begin{split}
  \int_{0}^{\infty}\!\mathrm{d}\omega\,
  \frac{W(\omega)}{\bar\omega+\omega}
  &=\ln\!\Bigl(1+\frac{1}{z}\Bigr)
  +\int_{0}^{1}\!\mathrm{d}t\,\frac{t^{2}}{1+zt}\\
  &=\ln\!\Bigl(1+\frac{1}{z}\Bigr)
  +\sum_{j=0}^{\infty}\frac{(-z)^{j}}{j+3},
  \end{split}
  \label{eq:W_integral}
\end{equation}
with $z=\bar\omega/\omega_{c}=\bar k/k_{c}$. Since $z\ll1$, the second
integral has been expanded in powers of $z$; its leading term $1/3$ is the
contribution of the weighted tail above $\omega_{c}$. Together with
Eqs.~\eqref{eq:bethe_exact} and \eqref{eq:sumrule} one obtains for S states
\begin{equation}
  \Delta E_{n}^{W}=\frac{4\alpha^{5}mc^{2}}{3\pi n^{3}}
  \Bigl[\ln\!\Bigl(1+\frac{k_{c}}{\bar k_{n}}\Bigr)
  +\frac13-\frac{z}{4}+\frac{z^{2}}{5}-\dots\Bigr].
  \label{eq:bethe_W}
\end{equation}
For $z\ll1$ the bracket reduces to $\ln(k_{c}/\bar k_{n})+\tfrac13$; the terms
of order $z$ are below $10^{-3}$ in all cases considered here.

The logarithm in Eq.~\eqref{eq:bethe_W} collects photon frequencies from
$\bar\omega_{n}$ up to $\omega_{c}$. Its ultraviolet end is not an atomic
property: at $k\gg\bar k_{n}$ the photon resolves the electron on length
scales far below the atomic size, down to the Compton wavelength. In the
object-relative picture, the localized object that sets the crossover in this
region is the electron, localized on its Compton wavelength. We therefore take
\begin{equation}
  k_{c}=k_{C}=\frac{mc}{\hbar},
  \qquad \omega_{c}=\frac{2\pi c}{\lambda_{C}}=\frac{mc^{2}}{\hbar}
  \label{eq:omega_c_compton}
\end{equation}
as a representative value; the geometric factor relating $k_{c}$ to the
Compton scale depends on the structure of the electron and is not fixed here.
This is also the scale at which the nonrelativistic description of the
electron underlying Eq.~\eqref{eq:bethe_exact} ceases to be valid.

With $\bar k_{2S}=k_{C}/2257$ and $n=2$, Eqs.~\eqref{eq:bethe_W} and
\eqref{eq:omega_c_compton} give
\begin{equation}
  \Delta E_{2S}^{W}=4.519\times10^{-6}\,\mathrm{eV}
  \quad(1092.7\,\mathrm{MHz}),
  \label{eq:lamb_result}
\end{equation}
which is $3.3\,\%$ above the measured value
$\Delta E_{\mathrm{Lamb,exp}}=4.3749\times10^{-6}\,\mathrm{eV}$. With Bethe's
original mean excitation energy of $17.8\,\mathrm{Ry}$ one obtains
$4.481\times10^{-6}\,\mathrm{eV}$ ($+2.4\,\%$).

Conversely, one may ask how far the crossover has to lie from the Compton
scale. Reproducing the measured value with Eq.~\eqref{eq:bethe_W} requires
\begin{equation}
  k_{c}=0.773\,k_{C}
  \qquad(0.827\,k_{C}\ \text{with}\ 17.8\,\mathrm{Ry}),
  \label{eq:kc_fit}
\end{equation}
i.e.\ a crossover wavelength $2\pi/k_{c}=1.29\,\lambda_{C}$. The sensitivity
is logarithmic: a change of $k_{c}$ by a factor $e$ shifts the result by
$135.6\,\mathrm{MHz}$, i.e.\ by $13\,\%$ of the Lamb shift.

This agreement should not be overstated, for three reasons. First, the Bethe
estimate contains neither the vacuum polarization (about $-27\,\mathrm{MHz}$
for 2S) nor the anomalous-moment contribution (about $+68\,\mathrm{MHz}$ to
the 2S--2P splitting) nor the small low-energy shift of the 2P level (about
$+4\,\mathrm{MHz}$). Agreement at the level of a few percent is what such an
estimate can deliver. Second, the Bethe logarithm drops out when the weighted
expression is compared with the relativistic QED result, since both contain
$\ln(\,\cdot\,/\bar k_{n})$. The comparison therefore constrains only the
ratio $k_{c}/k_{C}$, through the constant terms: matching
Eq.~\eqref{eq:bethe_W} to the leading-order QED self-energy of the 2S level
gives $k_{c}=0.824\,k_{C}$. The estimate thus shows that a crossover at the Compton
scale is consistent with the Lamb shift; it does not determine $k_{c}$ beyond
a factor of order unity. Third, when the sum over intermediate states in
Eq.~\eqref{eq:bethe_exact} is carried out explicitly instead of through the
mean frequency $\bar\omega_{n}$, the bracket in Eq.~\eqref{eq:bethe_W} changes
by $0.035$ at $k_{c}=k_{C}$ ($0.4\,\%$). This correction comes from continuum
states with excitation frequencies comparable to $mc^{2}/\hbar$, where the
nonrelativistic dipole treatment is no longer valid anyway.

The photon emitted and reabsorbed by the bound electron is an internal line
and therefore, according to Sec.~\ref{sec:hypothesis}, a virtual photon. In
Bethe's time-ordered treatment it appears as an intermediate quantum on the
light cone, and its virtuality is carried by the energy non-conservation of
the intermediate state; the frequency weighting used here is the natural form
of the weighting in that treatment. A covariant evaluation with
$W(K_{E})$ coincides with it below $k_{c}$ but differs near the crossover,
i.e.\ in the constant terms of Eq.~\eqref{eq:bethe_W}. It may therefore shift
the fitted $k_{c}$ by a factor of order unity, but not away from the Compton
scale.

The relativistic high-energy part of the Lamb shift is built from the
free-electron form factors $F_{1}'(0)$ and $F_{2}(0)$. A photon weighting with
$k_{c}\approx k_{C}$ acts there as well; this is the subject of the following
subsection.

\subsection{Form factors of the free electron: anomalous magnetic moment and relativistic part of the Lamb shift}
\label{sec:formfactors}

The anomalous magnetic moment provides a compact relativistic one-loop test
case for the proposed effective mode structure. For the internal photon line,
the weighting depends on the magnitude $K_{E}$ of the Euclidean photon
momentum (Sec.~\ref{sec:hypothesis}); the electron enters only through the
characteristic scale, which is taken near the electron Compton scale as in
Sec.~\ref{sec:bethe}.

The modified effective mode structure enters through the internal photon
contribution. In Feynman gauge, the photon propagator is
\begin{equation}
D_{\mu\nu}(k)=
\frac{-i\mu_{0}\hbar c^{2}}{(\hbar c)^{2}k^{2}+i\epsilon}\,
g_{\mu\nu},
\label{eq:photon_propagator}
\end{equation}
while the fermion propagator is
\begin{equation}
S(p-\hbar k)=
\frac{\slashed p-\hbar\slashed k+mc}{(p-\hbar k)^{2}c^{2}-m^{2}c^{4}+i\epsilon}.
\end{equation}
The one-loop self-energy is then schematically of the form
\begin{equation}
c\,\Sigma(p)
\sim
ic\int\frac{d^{4}k}{(2\pi)^{4}}\,
(-ie\gamma^{\mu})\,S(p-\hbar k)\,(-ie\gamma^{\nu})\,D_{\mu\nu}(k),
\end{equation}
with the on-shell subtraction fixing the physical mass and the residue of the
propagator,
\begin{equation}
\Sigma_{R}(p)=\Sigma(p)+\Sigma_{\mathrm{OS}}(p).
\end{equation}
As discussed in Sec.~\ref{sec:bethe}, this subtraction remains necessary but
is finite with the weighting.

After Wick rotation, the effective mode density can be inserted into the
Euclidean radial integral over the photon momentum. The Pauli form factor then
takes the form
\begin{equation}
F_{2}^{\mathrm{phys}}(q^{2})
=
\int_{0}^{\infty} dK_{E}\,
\frac{i2K_{E}^{3}}{(4\pi)^{2}}\,
W(K_{E})\,
F_{2}(K_{E},p,q),
\label{g2:F2_photon_radial}
\end{equation}
where $q=p'-p$ is the external momentum transfer, $K_{E}$ is the magnitude of
the Euclidean photon momentum, and $F_{2}(K_{E},p,q)$ denotes the angular
average of the integrand projected onto the $\sigma^{\mu\nu}q_{\nu}$
structure. Since the weighting acts on the photon line, this representation,
and not the Feynman-parametrized one below, is the appropriate one for the
weighted case.

For $W=1$ the standard reduction applies. After Feynman parametrization, Wick
rotation of the shifted loop momentum, and projection onto the
$\sigma^{\mu\nu}q_{\nu}$ structure, one obtains
\begin{equation}
\begin{aligned}
F_{2}(q^{2})
= &
\frac{\alpha}{\pi}
\int_{0}^{1}dx
\int_{0}^{1-x}dy\,
(1-x-y)(x+y)\\
&\times\int_{0}^{\infty} dK_{E}\,
\frac{4m^{2}c^{4}\,(\hbar c)^{4}K_{E}^{3}}
{\bigl(K_{E}^{2}(\hbar c)^{2}+\Delta(x,y,q^{2})\bigr)^{3}},
\end{aligned}
\label{g2:F2_kernel}
\end{equation}
with
\begin{equation}
\Delta(x,y,q^{2})
=
m^{2}c^{4}(x+y)^{2}-xy\,q^{2}c^{2}.
\end{equation}
Here $K_{E}$ is the shifted loop momentum, not the photon momentum. The
integrand falls as $K_{E}^{-3}$, so the integral is ultraviolet finite without
any weighting. With $\int_{0}^{\infty}dK_{E}\,(\hbar c)^{4}K_{E}^{3}/
(K_{E}^{2}(\hbar c)^{2}+\Delta)^{3}=1/(4\Delta)$ and
$\Delta(x,y,0)=m^{2}c^{4}(x+y)^{2}$,
\begin{equation}
F_{2}(0)=\frac{\alpha}{\pi}\int_{0}^{1}dx\int_{0}^{1-x}dy\,
\frac{1-x-y}{x+y}=\frac{\alpha}{2\pi},
\label{g2:schwinger}
\end{equation}
the Schwinger result~\cite{Schwinger1948b}.

The kernel must not be simplified by assuming $K_{E}\ll k_{C}$ in the
denominator. The integral collects most of its value around $K_{E}\approx
k_{C}$, and replacing one factor $K_{E}^{2}(\hbar c)^{2}+\Delta$ by $\Delta$
would turn the convergent kernel into a logarithmically divergent one.

The finiteness of Eq.~\eqref{g2:F2_kernel} has a structural reason. The
operator $\sigma^{\mu\nu}q_{\nu}$ connects opposite chiralities, so $F_{2}$
requires a mass insertion on the internal electron line, which replaces two
powers of loop momentum in the numerator by the factor $m^{2}c^{4}$.
Correspondingly, QED contains no Pauli term (it would be of mass dimension
five, so that its coefficient would carry the dimension of a length) into which a divergence could be absorbed, and $F_{2}(0)$ is a finite
prediction rather than a renormalized parameter. The anomalous magnetic
moment therefore does not test whether the weighting removes a divergence; it
tests how strongly the weighting may act at the scale at which $F_{2}$
collects its value.

With the piecewise weighting of Eq.~\eqref{eq:W_piecewise_KE}
applied to the photon momentum in Eq.~\eqref{g2:F2_photon_radial}, a direct
Euclidean evaluation gives $F_{2}^{\mathrm{phys}}(0)=R(k_{c})\,\alpha/2\pi$
with
\begin{equation}
1-R(k_{c})\simeq
\begin{cases}
8.6\times10^{-2}, & k_{c}=k_{C},\\
3.9\times10^{-3}, & k_{c}=3\,k_{C},\\
4\times10^{-5}, & k_{c}=10\,k_{C},
\end{cases}
\label{g2:F2_deficit}
\end{equation}
and $1-R\approx0.45\,(k_{C}/k_{c})^{4}$ for $k_{c}\gg k_{C}$. The fourth power
reflects that the piecewise profile is exactly unity below $k_{c}$, so that
only its ultraviolet tail acts. With the measured value
$a_{e}=1.15965218059(13)\times10^{-3}$~\cite{Fan2023}, a photon weighting
alone requires $k_{c}\gtrsim250\,k_{C}$. This is in tension with the
object-relative expectation $k_{c}\approx k_{C}$ and with the Lamb-shift
estimate of Sec.~\ref{sec:bethe}.

The same weighting also changes the slope of the Dirac form factor. Writing
$F_{1}(q^{2})\simeq1-\tfrac16\langle r^{2}\rangle Q^{2}$ for spacelike
momentum transfer, $Q^{2}=-q^{2}/\hbar^{2}$, the weighting at $k_{c}=k_{C}$
gives
\begin{equation}
\delta\langle r^{2}\rangle\simeq-1.1\times10^{-3}\,\bar\lambda_{C}^{2},
\qquad \bar\lambda_{C}=\frac{\hbar}{mc},
\label{g2:F1_slope}
\end{equation}
which enters the relativistic part of the Lamb shift and lowers the 2S level
by about $33\,\mathrm{MHz}$. Charge renormalization cannot absorb this. It
fixes only $F_{1}(0)=1$, which is preserved here by the Ward identity
(Sec.~\ref{sec:ward}), and it is carried by the vacuum polarization, which
contains no internal photon line and is not treated here. Like $F_{2}(0)$, the
slope of $F_{1}$ has no counterterm in QED (a charge-radius term would be of
mass dimension six, with a coefficient of the dimension of a length squared) and is a finite prediction.

If the weighting of the photon line is to sit at $k_{c}\approx k_{C}$, the
changes in Eqs.~\eqref{g2:F2_deficit} and \eqref{g2:F1_slope} must be
compensated by additional structure on the electron line at the same scale.
The size of the required structure can be stated. Parametrizing the electron
propagator Ward-consistently as $S(p)=[\slashed{p}-M(p^{2})c]^{-1}$ with
$M(m^{2}c^{2})=m$ and using the Ball--Chiu vertex~\cite{BallChiu1980}, one
finds on shell, exactly and already at tree level,
\begin{equation}
F_{1}(0)=1,\qquad
F_{2}^{\mathrm{tree}}=\frac{\eta}{1-\eta},
\label{g2:F2_tree}
\end{equation}
with the dimensionless slope
$\eta=2mc^{2}\,dM/dp^{2}\,|_{p^{2}=m^{2}c^{2}}$.
A slope of the mass function at the mass shell acts like a Pauli term and
leaves the Dirac form factor unchanged. The deficit of $F_{2}(0)$ in
Eq.~\eqref{g2:F2_deficit} is compensated by
\begin{equation}
\eta\simeq1.0\times10^{-4},
\label{g2:slope_needed}
\end{equation}
i.e.\ by a relative change of the mass function of order $10^{-4}$ over
$|p^{2}-m^{2}c^{2}|\sim m^{2}c^{2}$. The change of $F_{1}$ requires in
addition a charge-radius-type contribution
$\langle r^{2}\rangle\simeq+1.1\times10^{-3}\,\bar\lambda_{C}^{2}$. A mass
function with vanishing slope at the mass shell that grows above $k_{C}$ can
also restore $F_{2}(0)$ through the loop, but it enlarges the change of
$F_{1}$ (to about $-120\,\mathrm{MHz}$ in the 2S level) and is disfavoured.

In this form, the anomalous magnetic moment does not determine $k_{c}$.
Together with the Lamb shift, it quantifies what a photon weighting at the
Compton scale requires from the electron line.

\subsection{Open questions regarding the electron line}
\label{sec:open}

The results of Sec.~\ref{sec:formfactors} leave several questions open, all of
which concern the electron line rather than the photon line.

First, the required structure cannot come from the weighted one-loop
self-energy: with the same $W$ in self-energy and vertex, the modified mass
renormalization is already contained in $R(k_{C})$. It would have to be a
momentum dependence of the electron line itself, i.e.\ of the bare mass
$m_{0}$, which standard QED takes to be a constant. Such a structure is
meaningful only because the mass subtraction has become finite
(Sec.~\ref{sec:bethe}); in standard QED the bare mass is divergent and carries
no physical content of its own.

Second, to be compatible with $a_{e}$ and the Lamb shift within their
uncertainties, the slope $\eta$ and the charge-radius-type contribution would
have to be fixed to about $10^{-8}$ relative. A fit cannot supply this; it
requires a derivation of the electron-line structure from the same physics
that produces the photon weighting.

Third, the mode weighting derived here applies to photon modes. The one-loop
vacuum polarization consists of an electron loop and contains no internal
photon line. Whether an analogous structure exists for the virtual
electron--positron pairs of the vacuum polarization requires a treatment of the
mode structure of the electron itself; until then, the vacuum polarization is
renormalized as in standard QED. This question is sharply constrained by
experiment: in muonic hydrogen, the electron vacuum polarization is the
dominant contribution to the Lamb shift and is probed at momenta near the
electron Compton scale~\cite{Pohl2010,Antognini2013}. A change of the
electron loop at the percent level would shift the proton charge radius
extracted from these measurements far beyond its uncertainty. The electron-line
structure required above would also act inside the vacuum-polarization loop,
since the same electron propagators appear there, and must keep the vacuum
polarization transverse.

Fourth, the electron line cannot simply be given the photon weighting. Applying the
same $W$ to the electron propagators while leaving the vertex unchanged
violates the Ward--Takahashi identity, which relates the vertex to the inverse
electron propagator, and the vacuum polarization ceases to be transverse. With
$S\gamma^{\nu}S=-\partial S/\partial p_{\nu}$, its value at $q=0$ splits into
a surface term, which vanishes because $W$ falls off rapidly, and a term
proportional to the derivative of $W^{2}$, which does not. The latter is a
photon mass. For the piecewise profile one finds
$m_{\gamma}^{2}c^{4}\simeq(3\alpha/4\pi)(\hbar c\,k_{c})^{2}$ for
$k_{c}\gg k_{C}$ and $m_{\gamma}c^{2}\approx0.03\,mc^{2}\approx17\,
\mathrm{keV}$ for $k_{c}=k_{C}$, whereas the experimental bound is
$m_{\gamma}c^{2}<10^{-18}\,\mathrm{eV}$~\cite{PDG2024}. In standard QED, the
corresponding surface term is the quadratically divergent contribution that a
gauge-invariant regularization sets to zero; with the weighting, the
cancellation fails because the contributing terms carry different powers of
$W$, as for the single-vertex contributions in linearized
gravity~\cite{Rembe2026}. If, instead, the weighting is implemented such that
the electron loop is actually suppressed above $k_{c}$, the vacuum
polarization loses its contribution to the running of the fine-structure
constant above that scale, since at a large momentum transfer $Q$ at least one
electron line carries a momentum of at least $Q/2$. The electron contribution
to the running up to the $Z$ mass,
$\Delta\alpha_{e}(M_{Z}^{2})=(\alpha/3\pi)[\ln(M_{Z}^{2}/m^{2})-5/3]
\approx0.0174$, would essentially disappear; this is more than two orders of
magnitude larger than the uncertainty of the hadronic contribution, which
already limits electroweak precision tests~\cite{PDG2024}. A gauge-covariant
implementation, with the covariant derivative in the argument of $W$, avoids
the photon mass, but the Ward--Takahashi identity then requires vertices that
grow like $W^{-1}$, together with additional two-photon vertices. The
weighting then modifies the coupling itself and is no longer a weighting of
the mode content, and whether the suppression survives is not guaranteed:
higher-covariant-derivative regularization is known to leave one-loop
divergences, which have to be removed by additional Pauli--Villars
fields~\cite{Slavnov1972}. The reason is structural. The electron carries the
charge, so that for the electron line the mode content and the coupling
cannot be separated, whereas for the photon line they can. The electron line
therefore requires a structure of its own.

Fifth, the geometric factor between $k_{c}$ and $k_{C}$ is not fixed. The
Lamb-shift estimate favours $k_{c}$ at or slightly below $k_{C}$,
Eq.~\eqref{eq:kc_fit}; a factor of $2$ or $\pi$ above $k_{C}$ would already be
in tension with it. A change of the electron-line structure would, however,
also shift the relativistic part of the Lamb shift and with it this estimate.

These points have a common origin, the mode structure of the electron itself.
They are therefore not independent corrections but requirements on a single
structure, which is left to future work.

\section{Discussion}

The present work explores the possibility that ultraviolet contributions of electromagnetic modes are not determined solely by the standard vacuum mode counting, but may be effectively weighted relative to a localized interaction and its characteristic scale. The proposal is motivated heuristically and is not presented as a derivation from a complete underlying theory. Its purpose is more limited: to test whether an object-relative ultraviolet weighting of internal electromagnetic modes can be formulated in a Lorentz-consistent way, be implemented without spoiling the Ward identities, and render one-loop QED contributions with internal photon lines ultraviolet finite, and to see what it implies for precisely measured observables.

A central conceptual point is that the proposed weighting is not introduced as a universal modification of the vacuum alone. Rather, it is tied to a localized interaction object and to the corresponding invariant characteristic scale $k_c$. For real photons, a dependence on $k^\mu k_\mu$ alone is not sufficient to capture the physically relevant crossover, since $k^\mu k_\mu=0$ on the light cone; they are weighted by the object-relative frequency $u\!\cdot\! k$. Virtual photons, in contrast, are short-lived fluctuations off the light cone. The distinction between the object-relative mode structure and the Lorentz-invariant amplitudes of the fluctuations leads to a weighting of internal photon lines by the invariant $k^\mu k_\mu$, i.e.\ by the Euclidean photon momentum, with the object entering only through $k_c$. This distinction is not only conceptually natural; a weighting of the spatial photon momentum in loops does not decouple and is excluded by the anomalous magnetic moment.

This also distinguishes the weighting from a Wilsonian cutoff. A cutoff of that kind is a computational device whose scale drops out of all renormalized observables. The scale $k_{c}$, in contrast, is attributed to the interacting object, differs between physical objects, and has observable consequences that no renormalization can absorb, as the form factors of Sec.~\ref{sec:formfactors}
show.

Within this framework, the proposal has several encouraging features. First, the asymptotic weighting of order $k_c^3/K_E^3$ compensates the leading four-dimensional phase-space growth, so that one-loop diagrams with internal photon lines become ultraviolet finite; the mass subtraction remains, but becomes a finite conversion between bare and observed mass. Second, the weighting can be represented by a gauge-invariant kinetic term of the photon field that leaves the vertex unchanged, so that the Ward--Takahashi identities hold to all orders. Third, the Casimir effect is unchanged as long as the characteristic scale lies far above the modes that generate the force, which for real metals is satisfied by a wide margin. Fourth, a Bethe-type estimate of the Lamb shift, evaluated without a lower cutoff, places the characteristic scale near the electron Compton wavenumber, $k_c=0.77\,k_C$, in agreement with the object-relative expectation. The estimate determines $k_c$ only up to a factor of order unity, since the sensitivity is logarithmic.

The relativistic test case, however, shows that a weighting of the photon line alone is not sufficient. At $k_c\approx k_C$, the anomalous magnetic moment is lowered by $8.6\,\%$ and the slope of the Dirac form factor changes by an amount that lowers the 2S level by about $33\,\mathrm{MHz}$. Both quantities are finite predictions without counterterm in QED, so that neither mass nor charge renormalization can absorb the change. A photon weighting alone would require $k_c\gtrsim250\,k_C$ for the anomalous moment, in contradiction to the Lamb-shift estimate. Reconciling both requires additional structure on the electron line at the same scale, a slope of the mass function of order $10^{-4}$ together with a charge-radius-type contribution, fixed to about $10^{-8}$ relative. Such a structure cannot be fitted; it has to be derived. Nor can it be obtained by applying the photon weighting to the electron line: since the electron carries the charge, this would generate a photon mass, remove the observed running of the fine-structure constant, or, in a gauge-covariant form, no longer be a weighting of the mode content.

Several limitations remain substantial. Whether a propagator falling faster than $1/k^2$ is compatible with unitarity, given that it requires a spectral density that is not everywhere positive, remains open, as does the question how the object dependence of $k_c$ is expressed at the level of the action. The equivalence of the time-ordered and the covariant treatment of the Lamb shift is not guaranteed once a weighting is introduced. The vacuum polarization, which contains no internal photon line, is not treated, and the extension to higher-loop graphs, overlapping divergences, and non-abelian gauge theories has not been carried out.

These limitations are substantial, but they do not remove the motivation for studying such an effective ansatz. If the basic physical intuition is correct---namely that sufficiently localized electromagnetic modes should not be treated as spectrally inert with respect to the localization conditions relevant to the interaction---then one may expect the standard ultraviolet mode counting to become effectively modified before a complete theory is available. The open questions identified here are not scattered: they all concern the electron line and point to a single missing ingredient, the mode structure of the electron itself.

The main conclusion of the discussion is therefore deliberately modest: an object-relative ultraviolet weighting of photon modes appears to be a conceptually viable and technically workable ansatz for diagrams with internal photon lines, and it places the characteristic scale near the electron Compton scale. At that scale, it cannot reproduce the form factors of the free electron by itself, and the size of the required electron-line structure is quantified. Whether such a structure follows from the same physics, and whether the idea can be embedded into a fully consistent relativistic quantum field-theoretic framework, remain the central open questions.

\section{Conclusion}

In this work, localized electromagnetic interactions have been studied in terms of an effective object-relative ultraviolet weighting of photon modes. The central idea is that a sufficiently localized field mode should not be treated as effectively unchanged with respect to the conditions defining that mode. This motivates an effective crossover structure which leaves the infrared sector unchanged up to a characteristic scale and suppresses ultraviolet contributions asymptotically by a factor of order $k_c^3/k^3$.

A key conceptual step is the distinction between the mode structure and the amplitudes of the fluctuations. The mode structure is object-relative, and real photons are weighted by the object-relative mode variable $u_{\mu} k^{\mu}$, the mode frequency measured in the rest frame of the localized interaction object. The fluctuation amplitudes follow the rules of special relativity, and virtual photons in loops are weighted by the invariant $k^\mu k_\mu$, with the object entering only through the scale $k_c$.

Within this framework, one-loop contributions with internal photon lines become ultraviolet finite. The weighting can be formulated as a gauge-invariant kinetic term of the photon field, so that the Ward--Takahashi identities are preserved.

Four examples were discussed. The Casimir effect is unchanged as long as the characteristic scale lies far above the modes that generate the force. A compact ultraviolet one-loop test shows directly that the weighting suppresses the ultraviolet sector above the characteristic scale while leaving the infrared part unchanged. A Bethe-type Lamb-shift estimate places the characteristic scale near the electron Compton scale. For the form factors of the free electron, however, a photon weighting at this scale lowers the anomalous magnetic moment by $8.6\,\%$ and changes the slope of the Dirac form factor by an amount corresponding to $33\,\mathrm{MHz}$ in the 2S level. Reconciling both with experiment requires additional structure on the electron line, whose size has been quantified.

Several important open problems remain: the derivation of this electron-line structure, the treatment of the vacuum polarization, the compatibility of the weighting with a K\"allen--Lehmann-type spectral representation, the expression of the object dependence of $k_c$ at the level of the action, and the extension beyond one loop.

The present work is therefore not a replacement for ultraviolet renormalization in quantum electrodynamics. Rather, it is a structured first exploration of whether part of the ultraviolet problem may be reformulated in terms of an physically motivated object-relative weighting of electromagnetic modes. In that limited but precise sense, the results obtained here provide a motivated starting point for further investigation, and they identify the mode structure of the electron as the next task.

\begin{acknowledgments}
Generative AI (Claude Opus 5.5, Anthropic) was used to check derivations and calculations, to carry out numerical evaluations, and to assist in editing the text. All results were reviewed and verified by the author, who remains fully responsible for the scientific content of this manuscript.
\end{acknowledgments}



\begin{thebibliography}{99}\label{sec:TeXbooks}%


\bibitem{Lamb1947}
W. Lamb, R. Retherford,
Fine Structure of the Hydrogen Atom by a Microwave Method,
Physical Review 72, 241 (1947).
%
\bibitem{Dirac1958}
P.A.M. Dirac, The Principles of Quantum Mechanics, 4th ed, Oxford University Press, Oxford, 1958.
%
\bibitem{Weisskopf1934}
V.Weißkopf,
Über die Selbstenergie des Elektrons,
Zeitschrift für Physik 89, 27 (1934).
%
\bibitem{Bethe1947}
H. A. Bethe,
The Electromagnetic Shift of Energy Levels,
Physical Review 72, 339 (1947).
%
\bibitem{DrakeSwainson1990} G. W. F. Drake and R. A. Swainson,
Bethe logarithms for hydrogen up to n=20, and approximations for two-electron atoms,
Phys. Rev. A 41, 1243 (1990).
%
\bibitem{KundL1949}
N. M. Kroll, W. Lamb, On the Self-Energy of a Bound Electron, Physical Review 75, 388 (1949).
%
\bibitem{Tomonaga1946}
S.-I. Tomonaga,
On a Relativistically Invariant Formulation of the Quantum Theory of Wave Fields, Progress of Theoretical Physics 1, 27 (1946).
%
\bibitem{Schwinger1948a}
J. Schwinger, Quantum Electrodynamics I: A Covariant Formulation, Physical Review 74, 1439 (1948).
%
\bibitem{Feynman1949a}
R. Feynman, Space-Time Approach to Quantum Electrodynamics, Physical Review 76, 769 (1949).
%
\bibitem{PundV1949}
W. Pauli, F. Villars, On the Invariant Regularization in Relativistic Quantum Theory, Reviews of Modern Physics 21, 434 (1949).
%
\bibitem{Feynman1949b}
R. Feynman, The Theory of Positrons, Physical Review 76, 749 (1949).
%
\bibitem{Dyson1949}
F. Dyson,
The Radiation Theories of Tomonaga, Schwinger, and Feynman, Physical Review 75, 486 (1949).
%
\bibitem{Dyson1950}
F. Dyson, The S Matrix in Quantum Electrodynamics, Physical Review 75, 1736 (1949).
%
\bibitem{Feynman1950}
R. Feynman, Mathematical Formulation of the Quantum Theory of Electromagnetic Interaction, Physical Review 80, 440 (1950).
%
\bibitem{Salam1951}
A. Salam, Overlapping Divergences and the S-Matrix, Physical Review 82, 217 (1951).
%
\bibitem{Dirac1978}
P. Dirac, The Inadequacies of Quantum Field Theory, in: Mathematical Foundations of Quantum Theory, ed. A. R. Marlow (Academic Press, New York, 1978), pp. 1--7.
%
\bibitem{Feynman1985}
R. Feynman, QED: The strange theory of light and matter, Universities Press, 1985.
%
\bibitem{BundD1964}
J.D. Bjorken, S.D. Drell, Relativistic Quantum Mechanics, Volume 1, McGraw-Hill, 1964.
%
\bibitem{KundK1997}
G. Köpp, J. Krüger, Einführung in die Quantenelektrodynamik, (Teubner Studienbücher Physik, Stuttgart, 1997).
%
\bibitem{Ryder1995}
L.H. Ryder, Quantum Field Theory, 2nd ed., Cambridge University Press, Cambridge, 1996.
%
\bibitem{Demtroeder2015}
  W. Demtröder, Laser-Spektroskopie: Grundlagen und Techniken, Ed. 3, Springer, Berlin, Heidelberg, 1993.
  %
\bibitem{Born2013}
  M. Born, E. Wolf, Principles of Optics, Ed. 7, Cambridge University Press, Cambridge, UK, 2013.
  %
\bibitem{Kaellen1952}
G. Källén, On the Definition of the Renormalization Constants in Quantum
Electrodynamics, Helvetica Physica Acta 25(4), 417-434 (1952).
%
\bibitem{Lehmann1954}
H. Lehmann, Über Eigenschaften von Ausbreitungsfunktionen und Renormierungskonstanten quantisierter Felder, Nuovo Cimento 11, 342–357 (1954). 
%
\bibitem{Collins2004}
J. Collins, A. Perez, D. Sudarsky, L. Urrutia, and H. Vucetich,
Lorentz Invariance and Quantum Gravity: An Additional Fine-Tuning Problem?,
Phys. Rev. Lett. 93, 191301 (2004).
%
\bibitem{Schwinger1948b}
J. Schwinger, On Quantum-Electrodynamics and the Magnetic Moment of the Electron, Physical Review 73, 416 (1948).
%
\bibitem{Ward1950}
J. C. Ward, An Identity in Quantum Electrodynamics, Physical Review 78, 182 (1950).
%
\bibitem{Takahashi1957}
Y. Takahashi, On the Generalized Ward Identity, Nuovo Cimento 6, 371 (1957).
%
\bibitem{Slavnov1972}
A. A. Slavnov,
Invariant regularization of gauge theories,
Theor. Math. Phys. 13, 1064 (1972).
%
\bibitem{Wick1954}
G. C. Wick,
Properties of Bethe--Salpeter Wave Functions,
Physical Review 96, 1124 (1954).
%
\bibitem{Fan2023} X. Fan, T. G. Myers, B. A. D. Sukra, and G. Gabrielse,
Measurement of the Electron Magnetic Moment,
Phys. Rev. Lett. 130, 071801 (2023).
%
\bibitem{BallChiu1980} J. S. Ball and T.-W. Chiu,
Analytic properties of the vertex function in gauge theories. I,
Phys. Rev. D 22, 2542 (1980).
%
\bibitem{Casimir1948}
  H. B. G. Casimir,
  On the Attraction Between Two Perfectly Conducting Plates,
  Proceedings of the Koninklijke Nederlandse Akademie van Wetenschappen 51,
  793 (1948).
%
\bibitem{Bordag2001} M. Bordag, U. Mohideen, and V. M. Mostepanenko,
New developments in the Casimir effect,
Phys. Rep. 353, 1 (2001).
%
\bibitem{Lamoreaux1997} S. K. Lamoreaux,
Demonstration of the Casimir Force in the 0.6 to 6\,$\mu$m Range,
Phys. Rev. Lett. 78, 5 (1997).
%
\bibitem{MohideenRoy1998} U. Mohideen and A. Roy,
Precision Measurement of the Casimir Force from $0.1$ to $0.9 \;  \mu$m,
Phys. Rev. Lett. 81, 4549 (1998).
%
\bibitem{PDG2024}
S. Navas et al. (Particle Data Group), Review of Particle Physics,
Phys. Rev. D 110, 030001 (2024).
%
\bibitem{Rembe2026}
C. Rembe, Self-Gravitation of Mode Quanta in a Causal Resonator: One-Loop
Finiteness in Linearized Quantum Gravity and the Emergence of the Dark-Energy
Scale, arXiv:2609.14650 (2026).
%
\bibitem{Pohl2010}
R. Pohl et al., The size of the proton, Nature 466, 213 (2010).
%
\bibitem{Antognini2013}
A. Antognini et al., Proton Structure from the Measurement of 2S-2P Transition Frequencies of Muonic Hydrogen, Science 339, 417 (2013).
%
\end{thebibliography}
\end{document}